%% file: aaskaii_template.tex
\documentclass[a4paper,11pt]{article}
\usepackage{aaskaiid}
\usepackage{orcidlink}
\usepackage{CJK}
\usepackage{amsmath,siunitx}

\usepackage[dvipsnames]{xcolor}
\definecolor{forestgreen}{rgb}{0.13, 0.55, 0.13}

\title{Substructures in Planet-Forming Disks with the SKAO}
\ShortTitle{Substructures in planet-forming disks}

\ShortName{Yinhao Wu and Jessica Speedie et al.} 

\author[1,2]{Yinhao Wu \begin{CJK*}{UTF8}{gbsn}(吴寅昊)\end{CJK*} \orcidlink{0000-0003-3728-8231}}
\affiliation[1]{Shanghai Astronomical Observatory, Chinese Academy of Sciences, Shanghai 200030, People's Republic of China}
\affiliation[2]{School of Physics and Astronomy, University of Leicester, Leicester LE1 7RH, UK}
\emailAdd{yhwu0130@gmail.com}

\author[3,4]{Jessica Speedie \orcidlink{0000-0003-3430-3889}}
\affiliation[3]{Department of Physics \& Astronomy, University of Victoria, Victoria, BC, V8P 5C2, Canada}
\affiliation[4]{Department of Earth, Atmospheric, and Planetary Sciences, Massachusetts Institute of Technology, Cambridge, MA 02139, USA}
\emailAdd{jspeedie@mit.edu}

\author[5,6,7]{Sebasti\'an P\'erez \orcidlink{0000-0003-2953-755X}}
\affiliation[5]{Departamento de F\'isica, Universidad de Santiago de Chile, Av.\,Victor
Jara 3493, Santiago, Chile}
\affiliation[6]{Millennium Nucleus on Young Exoplanets and their Moons (YEMS), Chile}
\affiliation[7]{Center for Interdisciplinary Research in Astrophysics Space Science (CIRAS), Universidad de Santiago, Chile}

\author[8]{John D. Ilee \orcidlink{0000-0003-1008-1142}}
\affiliation[8]{School of Physics and Astronomy, University of Leeds, Leeds, UK, LS2 9JT, UK}

\author[9]{Takahiro Ueda (植田高啓) \orcidlink{0000-0003-4902-222X}}
\affiliation[9]{National Astronomical Observatory of Japan, 2-21-1 Osawa, Mitaka, Tokyo 181-8588, Japan}

\author[10,11]{Claudia Toci \orcidlink{0000-0002-6958-4986}}
\affiliation[10]{European Southern Observatory, Karl-Schwarzschild-Strasse 2, 85748 Garching bei München, Germany}
\affiliation[11]{Departamento de Fisica aplicada III, ETSI Universidad de Sevilla,  Camino de los Descubrimientos, 41092 Sevilla, Spain}

\author[12,13]{Daniel J. Price \orcidlink{0000-0002-4716-4235}}
\affiliation[12]{School of Physics \& Astronomy, Monash University VIC 3800, Australia}
\affiliation[13]{Univ. Grenoble Alpes, CNRS, IPAG, 38000 Grenoble, France}

\author[14]{Asmita Bhandare \orcidlink{0000-0002-1197-3946}}
\affiliation[14]{University Observatory Munich, Ludwig-Maximilians-Universität München, Scheinerstr. 1, 81679 Munich, Germany}

\author[15]{Eleonora Bianchi \orcidlink{0000-0001-9249-7082}}
\affiliation[15]{INAF-Osservatorio Astrofisico di Arcetri, Largo E. Fermi 5, I-50125, Firenze, Italy}

\author[14, 16]{Tilman Birnstiel \orcidlink{0000-0002-1899-8783}}
\affiliation[16]{Exzellenzcluster ORIGINS, Boltzmannstr. 2, D-85748 Garching, Germany}

\author[8]{Richard A. Booth \orcidlink{0000-0002-0364-937X}}

\author[17]{Tyler L. Bourke \orcidlink{0000-0001-7491-0048}}
\affiliation[17]{SKA Observatory, Jodrell Bank, Lower Withington, Macclesfield, Cheshire SK11 9FT, UK}

\author[18]{Gemma Busquet \orcidlink{0000-0002-2189-6278}}
\affiliation[18]{Departament de Física Quàntica i Astrofísica (FQA), Institut de Ciències del Cosmos (ICCUB), Universitat de Barcelona, Institut d'Estudis Espacials de Catalunya (IEEC), Spain}

\author[19]{Simon Casassus \orcidlink{0000-0002-0433-9840}}
\affiliation[19]{Departamento de Astronomía, Universidad de Chile, Casilla 36-D, Santiago, Chile}

\author[20]{Yi-Xian Chen \begin{CJK*}{UTF8}{gbsn}(陈逸贤)\end{CJK*} \orcidlink{0000-0003-3792-2888}}
\affiliation[20]{Department of Astrophysical Sciences, Princeton University, Princeton, NJ 08544, USA}

\author[21]{Claudio Codella \orcidlink{0000-0003-1514-3074}}
\affiliation[21]{INAF-Istituto di Radioastronomia, Via Gobetti 101, I-40129, Bologna, Italy}

\author[13]{Nicol\'as Cuello \orcidlink{0000-0003-3713-8073}}

\author[22,3]{Ruobing Dong \begin{CJK*}{UTF8}{gbsn}(董若冰)\end{CJK*} \orcidlink{0000-0001-9290-7846}}
\affiliation[22]{Kavli Institute for Astronomy and Astrophysics, Peking University, 5 Yiheyuan Rd., Haidian District, Beijing 100871, People's Republic of China}

\author[15]{Antonio Garufi \orcidlink{0000-0002-4266-0643}}

\author[13]{Greta Guidi \orcidlink{0000-0002-7002-8928}}

\author[23]{Cassandra Hall \orcidlink{0000-0002-8138-0425}}
\affiliation[23]{Department of Physics and Astronomy, The University of Georgia, Athens, GA 30602, USA}

\author[24]{Haochang Jiang \begin{CJK*}{UTF8}{gbsn}(蒋昊昌)\end{CJK*} \orcidlink{0000-0003-2948-5614}}
\affiliation[24]{Max-Planck-Institut für Astronomie, Königstuhl 17, 69117 Heidelberg, Germany}

\author[25]{Izaskun Jim\'enez-Serra \orcidlink{0000-0003-4493-8714}}
\affiliation[25]{Centro de Astrobiología (CAB), CSIC-INTA, Ctra. de Ajalvir Km. 4, 28850, Torrejón de Ardoz, Madrid, Spain}

\author[26]{Hauyu Baobab Liu (呂浩宇) \orcidlink{0000-0003-2300-2626}}
\affiliation[26]{Department of Physics, National Sun Yat-Sen University, No. 70, Lien-Hai Road, Kaohsiung City 80424, Republic of China}

\author[27]{Mayank Narang \orcidlink{0000-0002-0554-1151}}
\affiliation[27]{Jet Propulsion Laboratory, California Institute of Technology, 4800 Oak Grove Drive, Pasadena, CA 91109, USA}

\author[28]{Elenia Pacetti \orcidlink{0000-0003-1096-7656}}
\affiliation[28]{INAF–Istituto di Astrofisica e Planetologia Spaziali (IAPS), Via del Fosso del Cavaliere 100, 00133 Rome, Italy}

\author[29]{Jaime Pineda \orcidlink{0000-0002-3972-1978}}
\affiliation[29]{Center for Astrochemical Studies, Max-Planck-Institut für Extraterrestrische Physik, Giessenbachstrasse 1, D-85748 Garching, Germany}

\author[30]{Paola Pinilla \orcidlink{0000-0001-8764-1780}}
\affiliation[30]{Mullard Space Science Laboratory, University College London, Holmbury St Mary, Dorking, Surrey RH5 6NT, UK}

\author[15]{Linda Podio \orcidlink{0000-0003-2733-5372}}

\author[31]{Danai Polychroni \orcidlink{0000-0002-7657-7418}}
\affiliation[31]{INAF-Turin Astrophysical Observatory, via Osservatorio 20, I-10025, Pino Torinese, Italy}

\author[15]{Giovanni Sabatini \orcidlink{0000-0002-6428-9806}}

\author[28]{Eugenio Schisano \orcidlink{0000-0003-1560-3958}}

\author[32]{Leonardo Testi \orcidlink{0000-0003-1859-3070}}
\affiliation[32]{Department of Physics and Astronomy, University of Bologna, 40127 Bologna, Italy}

\author[31]{Diego Turrini \orcidlink{0000-0002-1923-7740}}

\author[13]{Marion Villenave \orcidlink{0000-0002-8962-448X}}

\author[33]{David Wilner \orcidlink{0000-0003-1526-7587}}
\affiliation[33]{Center for Astrophysics $|$ Harvard \& Smithsonian, Cambridge, MA 02138, USA}

\abstract{Disks of gas and dust orbiting young stars are the arenas and material reservoirs for planet formation. 
Over the past decade, multiwavelength observations, from infrared to radio, have resolved the spatial distribution of hundreds of protoplanetary disks in nearby star-forming regions, revealing a diverse zoo of substructures. These substructures are morphological features such as rings, gaps, spirals, vortices, asymmetries, warps, or clumps that trace variations in density, temperature, or composition relative to an otherwise smooth distribution of gas and dust.
Many unknowns persist as to the origin of these substructures, 
their role in planet assembly, and their true properties. 
SKA-Mid Band 5b continuum observations, offering angular resolutions of $\sim$$\ang{;;0.05}$ ($\sim$$\ang{;;0.15}$) with AA4 (AA*) at $12.5$ GHz / $2.4$ cm, 
will enable new progress at this frontier. 
In this chapter, we outline the open questions in the field of disk substructure that SKA-Mid is uniquely poised to address, with a lens on dust thermal emission. 
}

\begin{document}
\begin{CJK*}{UTF8}{ipxm}

\include{journal-names}
\maketitle

\section{Introduction}
Until a few decades ago, the idea of detecting planets beyond our own Solar System seemed nearly unthinkable. Today, thanks to remarkable advances in instrumentation and analysis techniques, thousands of exoplanets have been discovered, unveiling a striking diversity of planetary systems across the Galaxy. As a result, understanding the origins of these worlds has become a central objective of contemporary astrophysics. 

The formation of planets is a complex, multi-stage evolutionary process. Stars are born in clustered regions within massive, dense regions of molecular hydrogen known as giant molecular clouds \citep{Lada2003}. The combination of gravity and large-scale turbulent motion in these clouds leads to localized collapses under self-gravity \citep{Larson1981,MacLow2004}. As material condenses, the clouds fragment into smaller and denser regions \citep{Hoyle1953}.
Protostars grow as material from the surrounding cloud or `envelope' accretes through an infalling flow. The initially high stellar multiplicity fraction drops with time as unstable multiples decay into stable hierarchical systems \citep[e.g.][]{Bate2003}.

During the formation of protostars, not all of the surrounding envelope material accretes directly onto the central object(s). Due to angular momentum conservation \citep{Pringle1981}, a portion of the infalling gas and dust settles into a flattened, rotating structure surrounding the young star, the so-called protoplanetary disk (PPD), which is the cradle of planets. The disk may form once, or even be destroyed and re-form multiple times due to ongoing infall events, ram pressure stripping, and tidal interaction with other protostars \citep[e.g.][]{Bate2018}.

In this chapter, we primarily focus on disks around Class II sources, which correspond to the evolutionary stage where the protostellar envelope has mostly dissipated and the disk is visible in the infrared. Most substructure observations to date have been conducted in this stage. These disks not only regulate the subsequent accretion onto the star but also play a pivotal role in the redistribution of mass and angular momentum within the system \citep[e.g.,][]{Armitage2010}. Composed primarily of cold gas and fine dust grains, Class II disks typically span from tens to several hundreds of astronomical units in radius. They are now widely recognised as the birthplaces of planets \citep[e.g., see review by][]{Manara-PPVII}. 
Within these disks, dust grains undergo collisional growth and aggregate into pebbles, which under favourable conditions can eventually form planetesimals and even fully formed planets \citep[reviewed in][]{Drkazkowska-PPVII,Birnstiel2024}. Therefore, understanding the formation and evolution of PPDs is key to uncovering the common origin of stars and planets \citep{McCrea1960}.

Although PPDs were first predicted through theoretical studies \citep{Lynden-Bell-Pringle1974,Cassen-Moosman1981,Ruden-Lin1986,Kenyon-Hartmann1987}, their properties have only been confirmed and explored in detail through observations. As an observation-driven field today, the most effective way to study PPDs is through direct imaging with telescopes. Fortunately, the rapid development of astronomical facilities over the past decade, particularly the emergence of powerful radio interferometers such as the Atacama Large Millimeter/submillimeter Array (ALMA) and the Karl G. Jansky Very Large Array (VLA), has allowed us to resolve these disks with unprecedented clarity. Arguably, the most transformative discovery in this field over the past decade, dating back to the unveiling of the HL Tau image in 2014 --- the same year as the last Square Kilometre Array (SKA) science book \citep{SKA-2015white-paper, SKA-2015white-paper-chapter} --- is the detection of substructures in PPDs \citep{ALMA2015}. 

Direct imaging observations have revealed that PPDs exhibit remarkable diversity in their structures \citep[see review by][]{Bae2023}, chemical compositions \citep[reviewed in][]{Oberg-etal.2023}, and evolutionary stages \citep[see review by][]{Williams-Cieza2011}. In this chapter, we focus on how the upcoming SKA can be utilised to observe substructures within PPDs, highlighting the unique advantages, capabilities, and indispensable role of SKA in this context. Other key aspects closely related to planet formation will be addressed in different chapters of this updated edition of the SKA Science Book. Large surveys of PPDs, and the emerging prospects for studying their properties in a statistical framework with the Square Kilometre Array Observatory (SKAO), will be discussed in detail in the "Demographics of planet-forming disks with the SKAO" chapter \citep{Garufi-etal.2026}. Complementarily, the opportunities and challenges of observing disks at centimeter (cm) wavelengths, where free-free emission tends to dominate in the innermost regions, are addressed in the "Ionized gas emission in protoplanetary disks with the SKAO" chapter \citep{Guidi-etal.2026}, with a focus on the caveats and diagnostic potential of long-wavelength observations. In addition, for the rapidly growing field of chemistry in PPDs, "Unveiling complex chemistry in planet-forming disks with the SKAO" chapter \citep{Podio-etal.2026} demonstrated how observations of emission lines from heavier molecules by SKA, including carbon chains and rings, as well as prebiotic molecules whose peak emission falls in the centimeter regime, can open up new avenues of research.


\section{What Are Substructures?}\label{sec: What Are Substructures?}
\begin{figure}
\centering
\includegraphics[width=\textwidth]{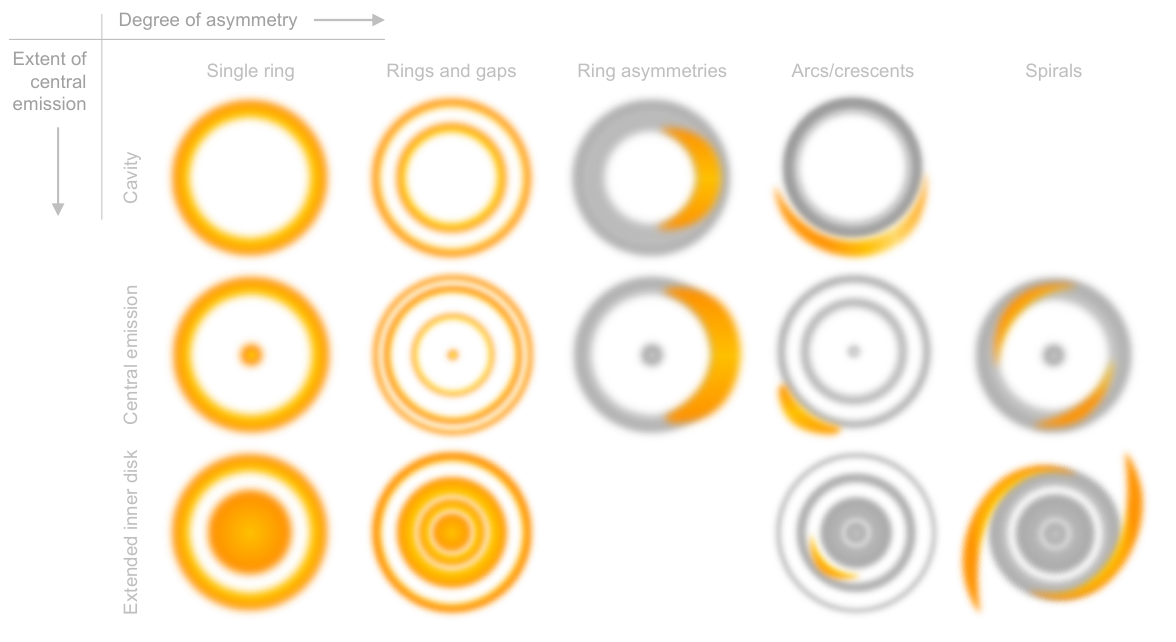}
\caption{
Schematic illustrations of the dust substructures observed in protoplanetary disks in mm continuum emission. 
The substructures are qualitatively organized in order of increasing asymmetry (to the right), and increasing extent of central emission (going downward). 
Within each disk illustration, yellow colour highlights the substructure that is labelled at the top of each column. 
Each illustration is based on a real disk, and blanks in the grid indicate no known examples. 
First row, left to right: 
J1604 \cite[231 GHz,][]{stadler2023-j1604}; 
LkCa 15, 
HD 34282, and 
HD 135344B \cite[332 GHz,][]{curone2025-exoALMA}.  
Second row: 
DM Tau  \cite[230 GHz,][]{Hashimoto2021};
HD 169142 \cite[232 GHz,][]{perez2019-hd169142}; 
AB Aur \cite[230 GHz,][]{tang2017-abaur}; 
HD 143006 \cite[240 GHz,][]{andrews2018-dsharp1}; 
HD100453 \cite[343 GHz,][]{rosotti2020-HD100453}. 
Third row: 
GW Lup \cite[240 GHz,][]{andrews2018-dsharp1}; 
HL Tau \cite[300 GHz,][]{ALMA2015}; 
blank; 
HD 163296 \cite[240 GHz,][]{andrews2018-dsharp1}; 
Elias 27 \cite[240 GHz,][]{huang2018-dsharp3}.
}\label{fig: substructures}
\end{figure}

Planet formation and its intricate relationship with PPDs have remained at the forefront of astrophysical research for decades. More precisely, what do substructures represent in the context of observations of disks? Substructures refer to morphological features discernible in resolved images of disks. These can currently be obtained in two main wavelengths regimes, in the millimeter (mm) and submillimeter (sub-mm) range, using facilities such as ALMA, NOEMA, and VLA \citep[e.g.,][]{Andrews2020}, and at optical to infrared wavelengths with space- and ground-based telescopes \citep[e.g.,][]{Benisty2023}, including the Hubble Space Telescope (HST), James Webb Space Telescope (JWST), Very Large Telescope (VLT), Gemini (North and South), and Subaru Telescope. 
As illustrated in Fig. \ref{fig: substructures}, these substructures include spiral arms, cavities, warps, crescents, and other azimuthal asymmetries, but most commonly, rings and gaps. 

The term {\em substructure} emerged to describe localized features {\em within} the broader structure of a protoplanetary disk. Before the advent of high-resolution facilities disks were generally modeled as smooth, axisymmetric structures. Although some disks were already known to deviate from this simple picture; most notably the so-called transitional disks \citep[e.g.,][]{Espaillat2014}, characterized by large inner cavities, and the lopsided or horseshoe-shaped disks revealed in ALMA Early Science results \citep[e.g.,][]{vanderMarel2013, Casassus2013}, these were still regarded as single, global structures rather than systems containing smaller-scale features. As higher-resolution imaging became available, observations began to reveal intricate fine patterns embedded within disks. The prefix “sub” thus highlights the emergence of these finer-scale, internal features within the overall disk structure.

It should be pointed out currently that most substructure studies have focused on Class II disks. Class 0 and Class I protostars represent the embedded phases of star formation, during which the central object is still deeply enshrouded in a collapsing envelope and the surrounding disk is often obscured at optical, near-infrared and even potentially millimetre wavelengths. However, recent high-resolution observations have begun to reveal disk substructures at earlier evolutionary stages \citep{Sheehan2018,SeguraCox2020,Maureira2024,Maureira2025}. In particular, surveys targeting these early-stage systems such as the ALMA large program Early Planet Formation in Embedded Disks (eDisk), have identified a few disks that already show ring-like features, spiral arms, or non-axisymmetric dust distributions \citep[][]{Ohashi-etal.2023eDisk}. Unfortunately, such detections remain relatively rare, and many early disks appear compact or only marginally resolved even with ALMA’s longest baselines \citep{Ohashi-etal.2023eDisk}. 

However, these results may reflect not the absence of substructures but the effect of high optical depths at (sub-)millimetre wavelengths that can obscure underlying features, as suggested by systems exhibiting only low-contrast substructures, e.g., IRS 63 \citep{SeguraCox2020}. Observations at longer wavelengths, where the dust emission becomes more optically thin, therefore provide a promising avenue to uncover such hidden structures. In this respect, the SKA will play a key role by enabling sensitive, high-resolution imaging of embedded disks in a regime less affected by optical depth effects. These findings suggest that substructure formation may either require time to develop or become more detectable as envelopes dissipate and the disk evolves into the Class II phase \citep[e.g.][]{CAMPOSII, Maureira-etal.2026}.

\subsection{Theoretical Framework}\label{sec: Theoretical Framework}

Here, we consider a range of mechanisms that can generate substructures in protoplanetary disks. Changes in dust properties near snowlines \citep{Zhang-etal.2015, Okuzumi-etal.2016}, intrinsic magnetohydrodynamic (MHD) instabilities \citep{Bai-Stone2014,Flock-etal.2015,Wu-etal.2023-winds,Wu2024,Su-Bai2025}, and gravitational instability \citep{dong16protostellar, speedie24} have been proposed as possible origins. By contrast, in many cases, substructures are thought to result from the interaction between embedded companions and their disks. 

Planets can carve dust-only or dust and gas gaps inside their parental disks \citep{Dipierro-etal2016}, with properties such as gap depth, gap separation, ring width, and ring thickness dependent on the mass ratio between the planet and the star \citep{Fung2017, DipierroLaibe17,Kanagawa2018,dong18doublegap, Bi2021, Bi2023}, as well as excite spirals and asymmetries  \citep{Fung2015, Zhu2015, Dong2015a}. 
Also, the presence of companion stars can induce prominent features such as spirals and asymmetries \citep[e.g., see reviews by][]{Zagaria-etal.2023,Cuello2025}, and close massive companions can lead to efficient truncation of the disk, producing large inner cavities or compact disks \citep{Artymowicz1994,Zurlo2023}.
These effects are best characterized by the mass ratio between the central star and its companion, which offers a dynamically meaningful criterion in distinguishing between stellar and planetary perturbers \citep[e.g.,][]{Dorazio16}. 


All these mechanisms are not mutually exclusive and may operate simultaneously in different regions of the same disk. Among the various mechanisms proposed, planet–disk interactions are particularly compelling due to their ability to produce sharply defined substructures \citep{Kley-Nelson2012} capable of slowing down dust radial drift. A massive planet embedded in the disk can gravitationally perturb the surrounding gas, carving a gap along its orbit \citep{Lin1980,Paardekooper-etal.2023}. In the dust component, this results in the accumulation of grains on either side of the gap, forming prominent ring-like features \citep{Ayliffe-etal2012,Pinilla2012}.

These features are often interpreted as dust traps. In such regions, grains are concentrated at pressure maxima in a size-dependent manner. Larger particles tend to be trapped more efficiently than smaller ones \citep[e.g.,][also see the Disk Demographics chapter by \cite{Guidi-etal.2026}]{Drkazkowska-PPVII,Birnstiel2024}, leading to narrower and more sharply defined rings when observed at longer wavelengths \citep{Speedie-etal.2022,Shi-etal.2024,Yang-etal.2025}. This behavior is a distinctive prediction of dust trapping and is not expected from alternative processes such as snowline-induced opacity variations. Multi-wavelength continuum observations provide strong support for this interpretation. In several systems, longer-wavelength data consistently reveal more compact ring structures that trace larger grains \citep{Carrasco-Gonzalez2019,Macias-etal.2019,Guidi2022,Doi-Kataoka2023,Liu-etal.2024,Sierra-etal.2024}.

In recent years, observational evidence supporting the planet–disk interaction hypothesis has grown significantly. Direct detections of forming planets have been reported inside disk cavities, most notably in the cavity of the PDS 70 disk \citep{Keppler2018,Close-etal.2025PDS70} and within the gap of the WISPIT 2 disk \citep{Close-etal.2025WISPIT,vanCapelleveen-etal.2025}, and candidates of spiral arm driving planets have also been found \citep[e.g.,][]{wagner23}. Indirect evidence has also emerged from deviations in molecular gas kinematics from Keplerian rotation \citep{Perez-etal-2015, Pinte-etal.2018,Perez-etal-2018, Casassus-etal.2019,Pinte-etal.2019,Pinte-etal.2020,Pinte-etal.2025}, which have been interpreted as signatures of gravitational perturbations by unseen planetary companions \citep[e.g.,][]{Pinte-etal.2023}, as well as spiral pattern speed measurements indicating companion origin \citep[e.g.,][]{ren18}. However, so far there have only been a small number of systems in which complementary spiral features have been found in mm continuum images, despite expectations that they should be detectable \citep[e.g.,][]{Speedie-etal.2022,Speedie-etal.2022b}. A plausible explanation is that high optical depths at mm wavelengths may hide the signal. Some examples supporting this idea are Elias 24 \citep{Carvalho2024} and PDS 66 \citep{Ribas-etal.2025}, which show significantly more structure at 3mm than 1.3 mm. 


\subsection{Observational Framework}\label{sec: Observational Framework}
Here we briefly introduce the observational framework to contextualize the signatures detected in disk images, followed by a discussion of how radio interferometry at longer wavelengths can uniquely constrain the nature, origin, and evolution of disk substructures. For a more comprehensive treatment of this topic, the reader is referred to the observational primer by \citet{Andrews2020}.

Resolving nearby protoplanetary disks in star-forming regions requires remarkably high sensitivity at sub-arcsecond angular resolution. To detect and characterize the finer substructures within these disks, angular resolutions on the order of tens of milliarcseconds (roughly 1-5 AU at 100 pc) are necessary. Such resolution can be achieved with 8-meter class telescopes operating in the optical and infrared, and with interferometers in the radio.

In terms of sensitivity, however, the challenge becomes more nuanced. Arguably, the most successful instruments in probing disk substructures to date have been ALMA and VLT/SPHERE, both of which have systematically delivered the sharpest and most detailed views.

The current observational landscape reveals that different regions of PPDs are best probed at different wavelengths and through distinct tracers. Scattered and polarized light in the near-infrared is particularly sensitive to small dust grains in the disk surface layers (VLT), while thermal continuum emission in the sub-mm (ALMA) and mm (VLA) regimes trace dust grains closer to the midplane. This has been a theoretical prediction for a very long time \citep{Weidenschilling1977} but is now confirmed with observations. This is easiest to do with edge-on disks \citep[e.g.,][]{Villenave-etal.2020,Villenave-etal.2022,Tazaki-etal.2025}, also many other works are starting to constrain settling at mm wavelengths on moderately inclined disks \citep{Pinte2016,Doi-Kataoka2021,Pizzati_2023}.

The most abundant molecule after H$_2$ is CO, and most of the advancements in characterizing substructures in the gas phase have been achieved through spectral line observations of CO isotopologues, particularly tracing the upper layers of the disk. Here, kinematic studies have been provided extremely helpful at recovering the velocity gradients revealing the underlying physics in planet-forming disks.

\subsection{Advances in Disk Observations}\label{Sec: Advances in Disk Observations}

\begin{figure}
\centering
\includegraphics[width=\textwidth]{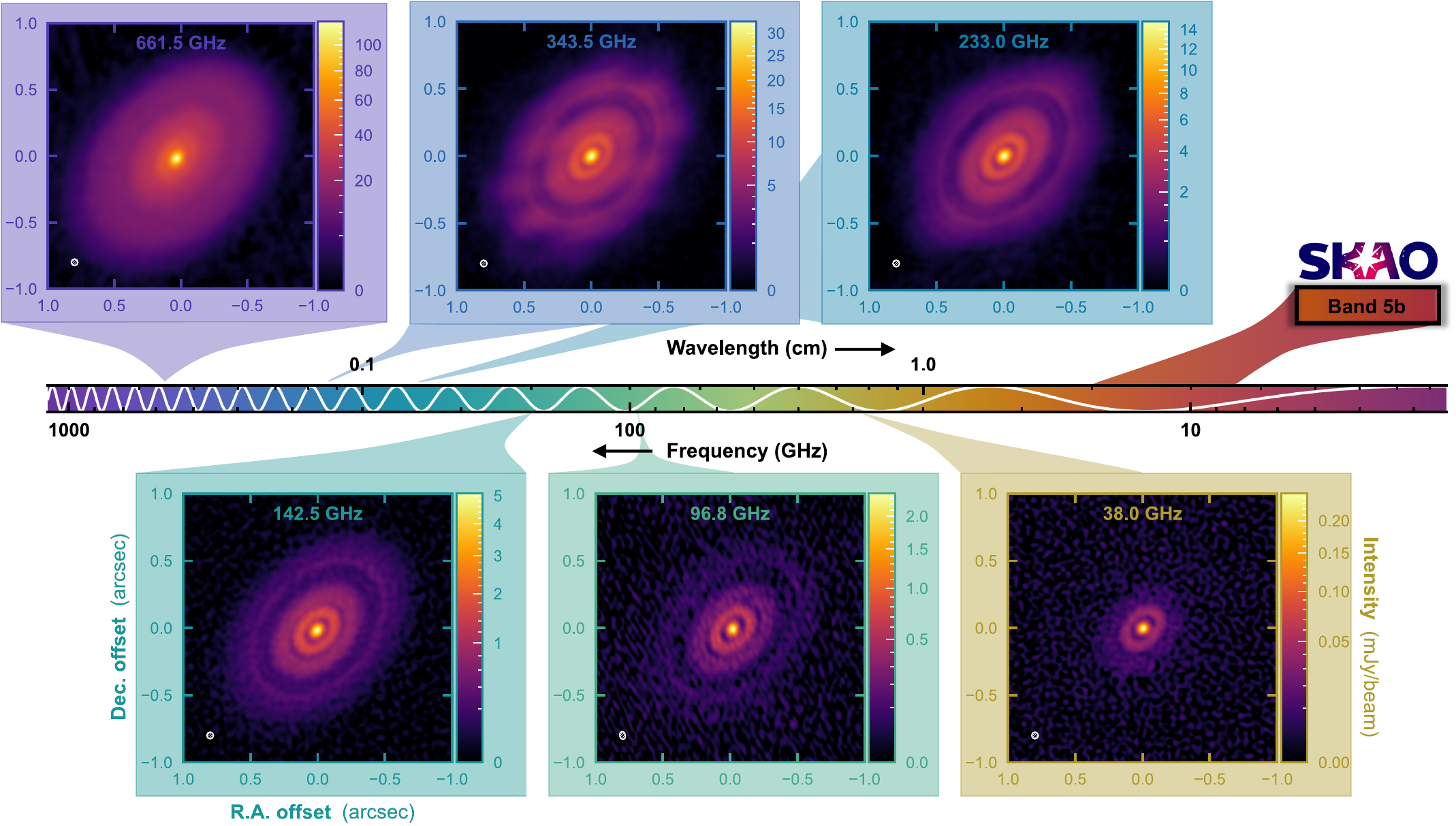}
\caption{
An overview of continuum observations, obtained in the last decade, spatially resolving the distribution of dust in the disk around HL Tau at (sub-)mm wavelengths. 
SKA-Mid will extend our view to sub-cm wavelengths that are presently unexplored, providing sensitivity to larger grains and optically thinner emission. 
This chapter focuses on continuum observations with Band 5b ($8.3 - 15.4$ GHz observing frequency range with $2\times 2.5$ GHz available bandwidth).
See Figures \ref{fig:hltau-AA*-AA4-radialprofile} \& \ref{fig:hltau-AA*-AA4} for our predicted SKA-Mid views of HL Tau. 
First row: 
ALMA Band 9 \cite[$0.45$ mm,][]{Guerra-Alvarado-etal.2024-HLTau0.45mm}; 
Band 7 \cite[$0.87$ mm,][]{ALMA2015}; 
Band 6 \cite[$1.3$ mm,][]{ALMA2015}. 
Second row: 
ALMA Band 4 \cite[$2.1$ mm,][]{Ueda-etal.2025}; 
Band 3 \cite[$3.1$ mm,][]{Ueda-etal.2025}; 
VLA Ka+Q \cite[$7.9$ mm,][]{Carrasco-Gonzalez2016, Carrasco-Gonzalez2019}.
}\label{fig:multiwavelength-hltau}
\end{figure}

Prior to the publication of \textit{Science with the SKA} book \citep{SKA-2015white-paper}, sub-arcsecond observations had already uncovered key signatures of large-scale structures in protoplanetary disks, particularly around Herbig Ae/Be stars. Among the most striking early results were the discoveries of prominent crescent-shaped dust concentrations near the edges of wide cavities in Oph IRS~48 \citep{vanderMarel2013} and HD~142527 \citep{Casassus2013}. These features, often referred to as dust traps, were interpreted as the result of gas pressure maxima concentrating mm-sized grains, offering some of the earliest indirect evidence of disk-planet or circumbinary interactions. Concurrently, the characterization of large inner cavities---typical of so-called transitional disks---was recognized as a hallmark of a critical evolutionary phase, potentially linked to planet formation. Additionally, gas observations of HD~142527 revealed non-Keplerian kinematics \citep{Rosenfeld2014, Casassus2015}, and when compared with shadows seen in scattered light imaging from the VLT, provided striking evidence that protoplanetary disks need not be confined to a single plane \citep{Marino2015}. Instead, strong misalignments between inner and outer disk regions are not only permitted by the physics, but are now known to be common in nature \citep{Ansdell2020,Bohn-etal.2022,Villenave-etal.2024}. While these early findings laid the foundation for subsequent surveys and modeling efforts, the precise origin and long-term stability of such asymmetric structures remain topics of active investigation \citep{vanderMarel2021}.

Then came the HL~Tau image in November 2014, a flagship result that demonstrated the transformative potential of ALMA's long-baseline capabilities. The image astonished the community and catalyzed a revolution in the theory and modeling of protoplanetary disks. Achieving a spatial resolution of approximately 4~au, the HL~Tau disk revealed up to seven bright rings and seven dark gaps \citep[the right one of first row in Fig. \ref{fig:multiwavelength-hltau},][]{ALMA2015}, tantalizing evidence that planet formation may already be underway in a disk less than a million years old \citep{Dong2015, Dipierro-etal.2015,Jin-etal.2016}.

Out of a sample of just over 100 PPDs observed at sufficiently high angular resolution, which typically means angular resolutions better than 0.05 to 0.1 arcseconds at sub-mm wavelengths, more than 50\% exhibit clear substructures such as rings, gaps, or asymmetries \citep{Long2018, vandermarel2019}. 
The remainder typically appears compact or featureless at current resolution limits. However, as observational capabilities continue to improve, particularly in terms of spatial resolution and sensitivity, some of the previously unresolved disks increasingly reveal substructures as well. Others, especially the compact disks, may remain smooth at mm wavelengths due to high optical depths \citep{Chung2024}. Observations at cm-wavelengths, where the optical depths are much lower, are essential to understanding the nature of the discs that appear smooth.

Although a wealth of observational evidence supports the idea that planets induce substructures in PPDs, a significant challenge remains in bridging the gap between mm-sized dust grains and kilometer (km)-scale planetesimals. Collisions between mm-sized particles frequently lead to fragmentation or bouncing, rather than the cohesive growth needed for planetesimal formation \citep[see review by][]{Blum-2008}. Moreover, solid particles in the disk experience aerodynamic drag from the surrounding gas, which drains their angular momentum and drives them inward toward the central star on rapid timescales \citep{Weidenschilling1977}. While mechanisms such as the streaming instability \citep[e.g.,][]{Youdin-2005, Johansen-2007, Li-etal.2019, Wu_Lin-2024} and pebble accretion \citep[e.g.,][]{Ormel-Klahr2010,Lambrechts2012,Johansen-Lambrechts2017} have been proposed to accelerate this growth process, the transition from mm-sized grains to cm-sized pebbles remains poorly understood and is critical to advancing our understanding of planet formation. 

\begin{figure}
\centering
\includegraphics[width=1\textwidth]{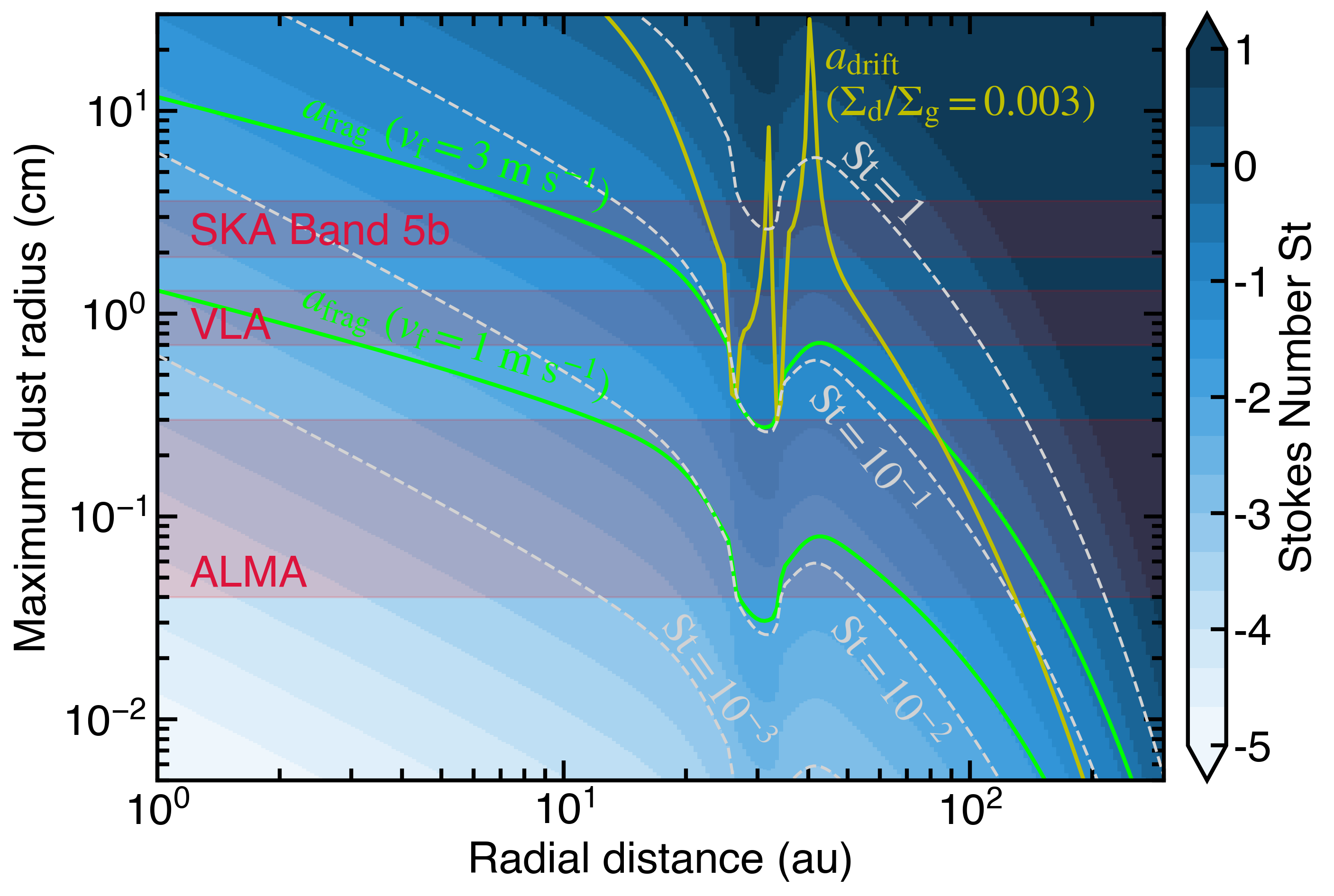}
\caption{
Maximum dust sizes
regulated by radial drift (yellow solid line) and collisional fragmentation (green solid lines) in protoplanetary disks \citep{Birnstiel2024}, overlaid with the typical dust sizes probed by ALMA, VLA, and SKA. The maximum sizes for drift and fragmentation are computed assuming a turbulence strength of $\alpha = 3\times10^{-4}$ and a gas surface density profile of $\Sigma_{\rm g}=1000\,(r/{\rm au})^{-1}\exp(-r/50~{\rm au})~{\rm g\,cm^{-2}}$ with including a gap structure expected for a Neptune-mass planet at 30 AU \citep{Kanagawa-etal.2017}.
}\label{fig:St_map}
\end{figure}

At mm wavelengths, dust opacity is primarily governed by mm-sized particles \citep{Draine-1984, Semenov-2003, Draine-2006}, making ALMA, which operates in the 0.3–3 mm range, highly effective at tracing this dust population. However, ALMA's sensitivity significantly diminishes for larger, cm-sized grains, limiting its ability to probe this crucial size regime.
Figure \ref{fig:St_map} presents a theoretical estimate of the maximum dust size attainable in protoplanetary disks.
The dust size in the disk is thought to be regulated either by radial drift ($a_{\rm drift}$) or by collisional fragmentation ($a_{\rm frag}$).
The maximum size determined by fragmentation strongly depends on the critical fragmentation velocity ($v_{\mathrm{f}}$; typically expected to be on the order of 1--10~m~s$^{-1}$), above which colliding dust aggregates break into smaller pieces.
Dust grains can grow beyond the millimeter regime probed by ALMA, and this observational challenge becomes more significant toward the inner regions of the disk.
The SKA will be able to directly trace cm-sized dust grains, providing a powerful observational handle on the largest solids and substantially alleviating the current limitations in testing dust evolution models.

Recent ALMA large programs, such as The Disk Substructures at High Angular Resolution Project \citep[DSHARP;][]{andrews2018-dsharp1}, Molecules with ALMA at Planet-forming Scales \citep[MAPS;][]{Oberg-etal.2021-MAPS-I}, exoALMA \citep{Teague-etal.2025-exoALMA-I}, and The ALMA Survey of Gas Evolution of PROtoplanetary Disks \citep[AGE-PRO;][]{Zhang-etal.2025-AGE-PRO-I}, have revealed that gas disks with CO emissions and dust disks traced at sub-mm/mm wavelengths can differ substantially in both spatial extent and morphological features. CO observations from the MAPS program, for example, show that gas disks often extend well beyond the continuum-emitting dust disk, with radii sometimes exceeding a few times than the sub-mm dust disk. Similarly, as mentioned in $\S$ \ref{sec: Theoretical Framework}, multi-wavelength observations \citep[e.g.,][]{Doi-Kataoka2023,Doi-etal.2024} have highlighted variations in radial structure between dust grains of different sizes, with large grains typically being more radially confined due to radial drift. These differences suggest that the disk morphology inferred from sub-mm observations alone may not fully capture the distribution of larger, mm- to cm-sized grains, which are more sensitive to local pressure structures and dynamical evolution. This difference becomes even more pronounced in edge-on systems \citep{Doi-Kataoka2023, Duchene-etal.2024, Tazaki-etal.2025}.

At present, however, VLA is the only operational facility capable of achieving sufficient sensitivity and angular resolution at cm wavelengths. A small number of disks have been successfully imaged with the VLA, including LkCa15 \citep{Isella2014}, HL Tau \citep{Carrasco-Gonzalez2016,Carrasco-Gonzalez2019}, HD 169142 \citep{Macias2017}, GM Aur \citep{Macias2018}, HD 163296 \citep{Guidi2022}, DM Tau \citep{Liu-etal.2024}, and MWC 480 \citep{Shi-etal.2026}, revealing substructures such as rings, gaps, and asymmetries at long wavelengths. These studies have provided important insights into dust growth and disk evolution, but typically require long integration times to reach the necessary sensitivity and resolution. As a result, its observational reach remains restricted to only the brightest handful of disks, limiting our ability to generalise findings across the broader disk population. This underscores the indispensable role of the SKA in the coming decade, as it will provide the capability to perform high-resolution imaging of PPDs at longer wavelengths. Compared to the VLA, the SKA will offer significantly improved sensitivity, broader frequency coverage, and a larger survey efficiency, which will enable observations of much fainter and more typical disks across different star-forming environments.

At present, however, VLA is the only operational facility capable of achieving sufficient sensitivity and angular resolution at cm wavelengths \citep{Garufi-etal.2025}. Yet, its observational reach is restricted to only the brightest handful of disks, limiting our ability to generalise findings across the broader disk population. This underscores the indispensable role of the SKA in the coming decade, as it will provide the capability to perform high-resolution imaging of PPDs at longer wavelengths. Compared to the VLA, the SKA will offer significantly improved sensitivity, broader frequency coverage, and a larger survey efficiency, which will enable observations of much fainter and more typical disks across different star-forming environments.


\subsection{Frontiers in Theory and Modeling}\label{sec: Frontiers in Theory and Modeling}
Since the publication of \textit{Science with the SKA} book \citep{SKA-2015white-paper} and the commissioning of ALMA, theoretical studies of planet formation and disk physics have seen substantial advances, driven in large part by the need to interpret increasingly detailed observations. Although the focus of this chapter is on substructures, we have also reviewed here several key theoretical developments over the past decade that are directly or indirectly related to them, and we hope that the next decade with the SKA will bring even more insights. One major development has been the refinement of dust evolution models that track the growth, fragmentation, radial drift, and vertical settling of grains over a wide size distribution \citep[e.g.,][]{Birnstiel-etal.2016,Stammler-etal.2019}. These models now routinely couple dust dynamics with gas disk evolution, enabling more realistic predictions of observable dust structures and their connection to underlying physical processes \citep[see review in ][]{Birnstiel2024}.

Simultaneously, hydrodynamical simulations incorporating embedded planets have achieved higher spatial and temporal resolution, allowing for robust modelling of planet-disk interactions across a range of disk viscosities, aspect ratios, and thermodynamic conditions \citep[e.g.,][]{Dong-etal.2017, Bae-etal.2017, Zhang-etal.2018-DSHARP-VII, Harter2020, Muley2021, Zhang2024}. These efforts have demonstrated that even low-mass planets can produce prominent rings, gaps, and vortices under the right conditions, particularly in low-viscosity environments. This has narrowed the gap between observed substructures and their theoretical counterparts, providing a framework for using disk features as indirect probes of young planets.

Historically, theoretical consensus regarding angular momentum transport and accretion in PPDs has been built around turbulence-driven accretion, i.e. the classical $\alpha$-model \citep{Shakura-Sunyaev1973}. However, the advent of ALMA has begun to challenge this paradigm. As discussed in $\S$ \ref{sec: Observational Framework} and $\S$ \ref{Sec: Advances in Disk Observations}, high-resolution and multi-wavelength observations now enable unprecedented probes of disk kinematics. Techniques such as line broadening analysis, spatially resolved gas kinematics, and statistical studies of disk evolution across large samples have been used to characterize turbulence levels \citep[see review by][]{Rosotti2023}. These studies increasingly suggest that many PPD systems, especially in their planet-forming regions, which exhibit turbulence levels significantly lower than previously expected. While the specific methodologies vary, and some gas tracers probe the upper disk layers rather than the midplane, the most recent constraints place the turbulent $\alpha$ parameter in the range of $10^{-4}\sim10^{-3}$ \citep{Flaherty-etal.2017,Flaherty-etal.2020}. If disks are indeed weakly turbulent or nearly inviscid, then a viable alternative mechanism must exist to account for how PPDs accrete and evolve over time. This carries profound implications, not only challenging the ubiquity of turbulence-driven accretion but also motivating the development of alternative models for disk evolution and stellar feeding.

Recent theoretical work has proposed that magnetically driven disk winds, or MHD disk winds \citep{Armitage-etal.2013,Bai-Stone2013,Bai2013,Suzuki-etal.2016,Bai-etal.2016,Lesur2021,Cui-Bai2021,Cui-Bai2022,Tabone-etal.2025}, may play a dominant role in removing angular momentum from the gas, enabling accretion at rates higher than those driven by disk turbulence. Moreover, the latest 2D hydro or 3D MHD numerical simulations suggest that the presence of MHD winds can significantly alter the emergence and morphology of disk substructures compared to purely laminar disks \citep{Elbakyan-etal.2022,Aoyama-Bai2023,Wu-etal.2023-winds,Wafflard-Fernandez-Lesur2023,Wu2024,Hu-etal.2025}. This raises the intriguing possibility that substructure morphology may offer a straightforward observational diagnostic for identifying wind-active systems.

Recent years have also seen an expansion of planet formation theory to include the role of disk substructures themselves. Pressure bumps, zonal flows, and vortices, which once considered consequences of planet formation, are now being studied as potential precursors that facilitate pebble concentration and gravitational collapse into planetesimals \citep{Surville-etal.2016, Taki-etal.2021}. 
In particular, the concept of "planet formation in dust traps" has gained traction, linking the spatial distribution of solids to the efficiency of core accretion and pebble accretion \citep{Bitsch-etal.2018, Jiang-Ormel2023, lau-drazkowska-2022}. These studies highlight how early disk conditions and substructure formation can regulate both the timing and location of planet formation.

In parallel, planet formation studies have highlighted the role of planetesimals in shaping the dust features in PPDs \citep[see also the Disk Demographics chapter by][]{Garufi-etal.2026}. In disks whose dust and gas are efficiently converted into planetesimals, as argued for the Solar System based on meteoritic data \citep{Scott2007,Lichtenberg2023,Sirono2025} --- forming planets will excite their surrounding planetesimal disk on eccentric and inclined orbits and trigger a chain of dust-producing collisions \citep{Turrini2012,Turrini2019,Bernabo2022,Sirono2025} akin to those responsible for the formation of debris disks at later stages in the life of planetary disks. These dynamical and collisional processes of the planetesimal disk imprint on the spatial distribution of cm-sized particles, injecting new dust and pebbles in disk regions that would otherwise be dust-depleted by the effects of dust traps, and creating dust sub-structures and rings in systems where multiple planets are forming \citep{Turrini2019,Bernabo2022}.

Another emerging frontier is the coupling of chemical and dynamical evolution in disk models. Time-dependent thermo-chemical codes now track how disk chemistry responds to evolving dust and gas distributions, which is crucial for interpreting molecular line emission and planet-forming environments \citep[][see also "Disk Chemistry" chapter by \cite{Podio-etal.2026}]{Eistrup2016,Eistrup2018,Booth2019,Pacetti2022,Pacetti2025}. 
Such models are increasingly important for connecting observable tracers (e.g., C/H ratios, CO emissions) with hidden processes like core accretion or atmospheric recycling.

Finally, the integration of machine learning and surrogate modeling techniques has opened new possibilities for forward modeling and statistical inference. Emulators trained on large grids of hydrodynamic or radiative transfer simulations can rapidly explore parameter spaces and constrain disk or planet properties from observed features \citep{Auddy-etal.2022,Mao-etal.2023,Mao-etal.2024,mao25, Ruzza-etal.2025}. These developments promise to transform the interpretation of high-resolution observations in the coming decade.

In sum, the post-2014 era has marked a transition from idealized, static disk models to dynamic, multi-physics frameworks that closely mirror the complexity revealed by modern observations. The next generation of theoretical work, especially when paired with the long-wavelength capabilities of the SKA, will be critical for connecting the observed diversity of disk structures to the underlying physics of planet formation.

\section{Simulated Observations \& Predictions for SKA-Mid Band 5b}\label{sec:synthetic-predictions}


We can directly demonstrate the capabilities of SKA Mid for imaging known disk substructures at centimetre wavelengths. 
Here we can be guided by models that have been well benchmarked against
high sensitivity and high angular resolution millimetre wavelength observations.\footnote{The models and data used in this chapter are publicly available: \href{https://doi.org/10.7910/DVN/CNBLEH}{https://doi.org/10.7910/DVN/CNBLEH}.} 

In this section, we demonstrate the science outcomes enabled by the SKA-Mid design baseline, Array Assembly 4 (AA4). 
The AA4 design baseline will consist of 133 15-m SKA dishes together with 64 13.5-m MeerKAT dishes, for a combined total of 197 dishes observing  simultaneously. 
The SKA dishes form AA4's longest baselines along three spiral arms (achieving a maximum baseline length of 159.6 km), while the MeerKAT dishes sit at the heart of AA4. 
Note that, in order to carry out observations with the full integrated assembly of 197 dishes in Band 5b, specifically, the 64 MeerKAT dishes will need to be retrofitted with receivers capable of operating in Band 5b. 
Here we consider the AA4 design baseline to be comprised of the 133 15-m SKA dishes alone. If or when MeerKAT is equipped with Band 5b receivers, we expect that AA4 will operate with a sensitivity roughly $1.4\times$ better than our predictions in this chapter represent.\footnote{Approximating the collecting area of the 64 13.5~m MeerKAT dishes as equivalent to that of 52 15~m dishes, and considering sensitivity scales as $\sqrt{N(N-1)}$, we obtain a sensitivity boost by a factor of $\approx$~$1.39$ for AA4, or by a factor of $\approx$~$1.65$ for AA*.}  


A secondary goal of this section is to document the extent to which scientific outcomes might be achieved by the end of staged delivery (AA*), chronologically sooner than AA4. 
The AA* array assembly is comprised of 80 15-m SKA dishes, all equipped with Band 5b receivers. 
The location of one particular dish, named SKA008, contributes the longest baselines to AA*, but introduces complexities for our image-plane method (see below) for predicting AA* imaging capabilities. 
For details, we refer the reader to Appendix $\S$\ref{sec:appendix}.  
In order to avoid these complexities, we consider AA* operating \textit{without} dish SKA008. Hereafter, all references to AA* should be understood to refer to the remaining 79 15-m dishes, with a maximum baseline of 36.0 km. 
As above, if the 64 MeerKAT dishes with Band 5b receivers are additionally included in AA*, we expect a sensitivity boost by a factor of approximately $1.7\times$ compared to the estimates adopted in this chapter.$^{1}$


The very large number of dishes in AA4 ---and, to a lesser but notable extent, in AA*--- will deliver exceptionally dense $uv$ coverage, marking a genuine step change relative to existing interferometric facilities. 
Here, we exploit this dense sampling of the $uv$ plane to construct SKA-Mid continuum images using an image-plane approach. 
We first generate model images of the expected 12.5 GHz continuum emission for a selection of known protoplanetary disks that are bright, spatially extended and structured. Our modelling of their 12.5 GHz emission is informed by their observed emission at ALMA and/or VLA frequencies. 
These model images are then convolved with a Gaussian kernel representing the synthesized beam, whose dimensions are found by fitting the point-spread function of $uv$-coverage simulated with the SKAO Observing Support Tool \texttt{ska\_ost\_array\_config} 
($\S$\ref{subsec:HLTau-with-AA*-AA4}) or directly taken from the SKAO Sensitivity Calculator version 2.1.1 ($\S$\ref{subsec:6-structured-disks-with-AA4}). 
Correlated noise is added with an rms provided by the SKAO Sensitivity Calculator. 
As this procedure implicitly assumes a fully sampled $uv$ plane, it does not capture effects associated with incomplete interferometric sampling or spatial filtering. We verified that, when integration times 
are sufficiently long, this image-plane method produces results comparable to full visibility-to-image simulations \cite[e.g.,][]{Ilee-etal.2020}. 

The remainder of this section is structured as follows. 
In $\S$\ref{subsec:HLTau-with-AA*-AA4}, we consider the HL Tau system for 
observations with AA* and AA4. 
Considering the revolutionary impact on the field made by ALMA Science Verification observations of the HL Tau disk ($\S$\ref{Sec: Advances in Disk Observations}), this system may be a conspicuous choice for early observations with SKA-Mid. 
Using this system as a comparison testbed, we demonstrate the improved capabilities AA4 will provide over AA*. 
In $\S$\ref{subsec:6-structured-disks-with-AA4}, we focus on the AA4 design baseline, and present simulated SKA-Mid images of a larger sample of structured disks, working with integration times up to a 1000-hour Key Science Project (KSP). 
Finally, in $\S$\ref{subsec:visibility-analyses} we describe the benefits of visibility-based analyses and advocate for the delivery of visibility data by SKAO.
\subsection{HL Tau: A Benchmark T Tauri Star}\label{subsec:HLTau-with-AA*-AA4}

As discussed in $\S$\ref{Sec: Advances in Disk Observations}, the HL Tau system represents a landmark in protoplanetary disk science. 
ALMA's first long-baseline images of this young 
T Tauri star revealed a striking system of concentric dust rings and gaps \citep{ALMA2015}, 
suggesting that planet formation may begin much earlier than previously anticipated, and prompting a new generation of high-angular resolution surveys that established substructure as common. 
Consequently, HL Tau now anchors both theoretical and observational work on planet formation, and boasts one of the most extensive, high-resolution, multi-wavelength data sets available for any protoplanetary disk (c.f. Figure \ref{fig:multiwavelength-hltau}). 

Recently, \citet{Ueda-etal.2025} presented a comprehensive analysis of HL Tau's dust disk by modelling its intensity profile observed at high angular resolution ($\ang{;;0.05}$ or $\sim$7 au) across six wavelengths (0.45–7.9 mm) with ALMA and the VLA. With this six-wavelength dataset (shown in Figure \ref{fig:multiwavelength-hltau}) and a Markov Chain Monte Carlo (MCMC) method, they constrained six key properties of the dust grain distribution, including temperature, surface density, maximum grain size, composition, filling factor, and size distribution. 
Their analysis could not, however, distinguish between amorphous-carbon-rich or organics-rich dust composition (a fraction captured in their model by a $f_{\rm AC}$ parameter), emphasizing the need for future observations at longer wavelengths where the dust is optically thin. 


Here, we employ the modeling results from \citet{Ueda-etal.2025} to generate a prediction for the radial intensity profile of the HL Tau disk at an observing wavelength of 2.4 cm, or 12.5 GHz (SKA-Mid Band 5b). Following $\S$5.5 of \citet{Ueda-etal.2025}, this model intensity profile is derived from the MCMC posterior probability distributions of the six constrained dust properties at each radial position within the disk. We optimistically assumed a dust composition rich in amorphous carbon, with $f_{\rm AC} = 0.75$-0.85. 
This choice has a significant impact on the predicted long-wavelength emission: if an organics-rich composition is adopted instead, the resulting centimeter intensity can be lower by a factor of $\sim$2--3. While the current data do not allow us to distinguish robustly between these compositions, previous disk population synthesis studies and a detailed disk modeling have suggested that an amorphous carbon-rich composition may be plausible \citep{Delussu-etal.2024,Zagaria-etal.2025}, which motivated our fiducial assumption.
From the posterior probability distributions, we estimated the 68\% confidence interval of the intensity at 12.5 GHz, and adopted the maximum values within this interval as the model intensity profile. 
We verified that this choice has only a modest effect compared with the adopted dust composition: using the minimum values within the same 68\% interval changes the predicted intensity by only $\sim$15\%.
This model radial intensity profile has a native resolution of $\ang{;;0.05}$ inherent to the observations from which it was derived, and forms the basis for the simulated SKA-Mid observations of HL Tau presented in the remainder of this subsection.


\begin{figure}
\centering
\includegraphics[width=\textwidth]{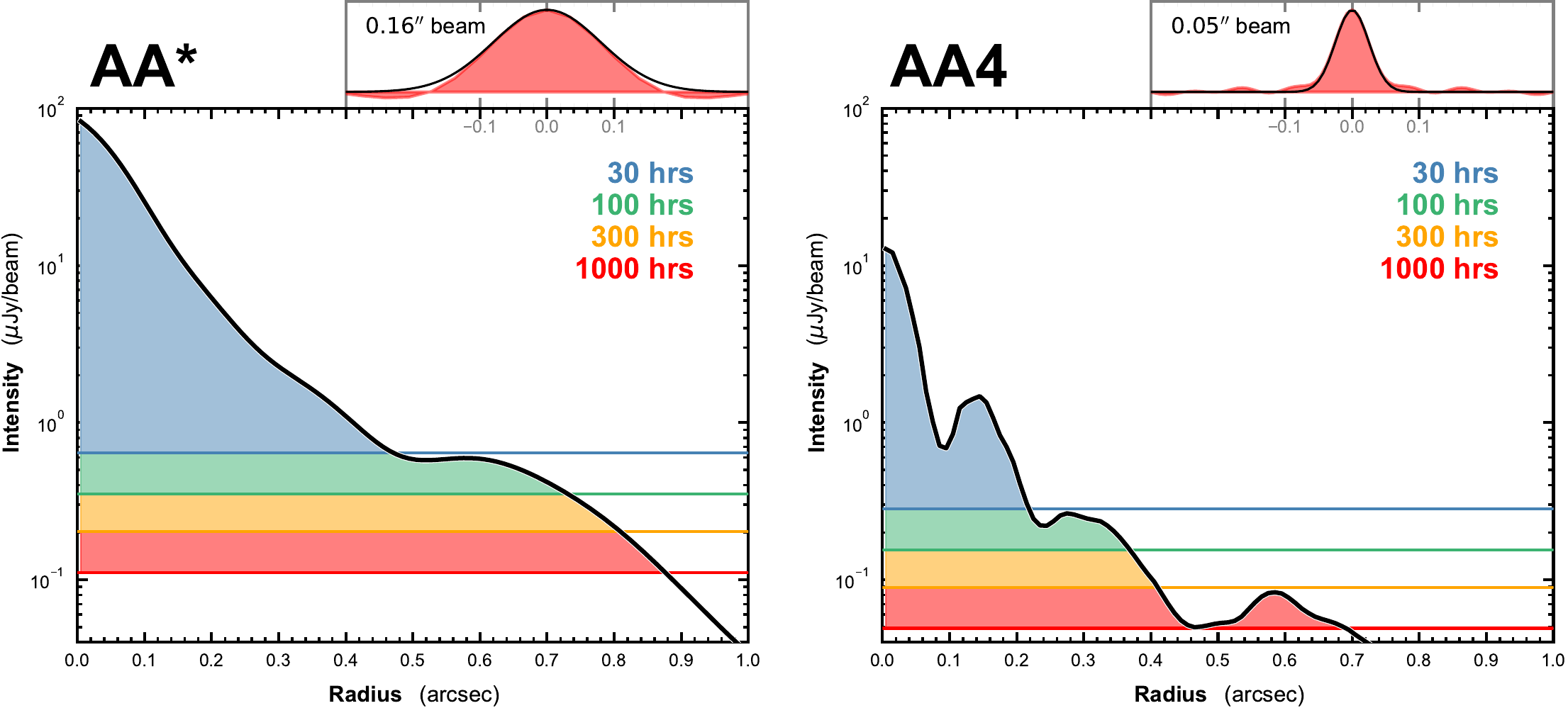}
\caption{
Predicted radial intensity profiles of the HL Tau disk at $12.5$~GHz (SKA-Mid Band 5b) with AA* (left panel) and AA4 (right panel). 
In each panel, the underlying model profile in $\mu$Jy~arcsec$^{-2}$ 
is scaled and spatially convolved to $\mu$Jy~beam$^{-1}$, where 
the beam 
is a Gaussian fit to the PSF 
cross section that is shown in the top right corner. 
The PSFs are generated from $uv$-coverage simulations, and weighted with a Briggs robust parameter of $-1$. 
Dish SKA008 is excluded from AA*. 
Horizontal lines show the rms noise or $1\sigma$ sensitivity achieved 
in 30~hours (blue), 100~hours (green), 300~hours (orange) and 1000~hours (red) of integration time. 
Synthetic images for the 1000~hour cases are shown in Figure \ref{fig:hltau-AA*-AA4}. 
}\label{fig:hltau-AA*-AA4-radialprofile}
\end{figure}

As a first assessment of how the angular scale of the HL Tau disk's radial substructure compares to the beam sizes accessible with SKA-Mid, and of how the predicted flux compares to the achievable sensitivity, 
Figure \ref{fig:hltau-AA*-AA4-radialprofile} shows 
the model radial intensity profile of the HL Tau disk at 12.5 GHz convolved to the angular resolution of AA* (79 15-m antennas) and AA4 (133 15-m antennas). 
The underlying model profile (in $\mu$Jy~arcsec$^{-2}$) is scaled and spatially convolved to $\mu$Jy~beam$^{-1}$. 
The beam solid angle is taken as the minor axis of a 2D Gaussian fit to the point-spread function (PSF) of the simulated $uv$-coverage, 
weighted with a Briggs robust parameter of $-1$. 
For comparison with the intensity profiles, we overlay horizontal lines at the $1\sigma$ continuum sensitivity estimated by the SKAO Sensitivity Calculator for increasing integration times. 
Our selection of integration times is meant to represent a range between a small program and Key Science Project. 
We find that detecting HL Tau's dust thermal emission with AA* requires modest integration times, while resolving its rings with AA4 requires deeper integrations.  



\begin{figure}
\centering
\includegraphics[width=\textwidth]{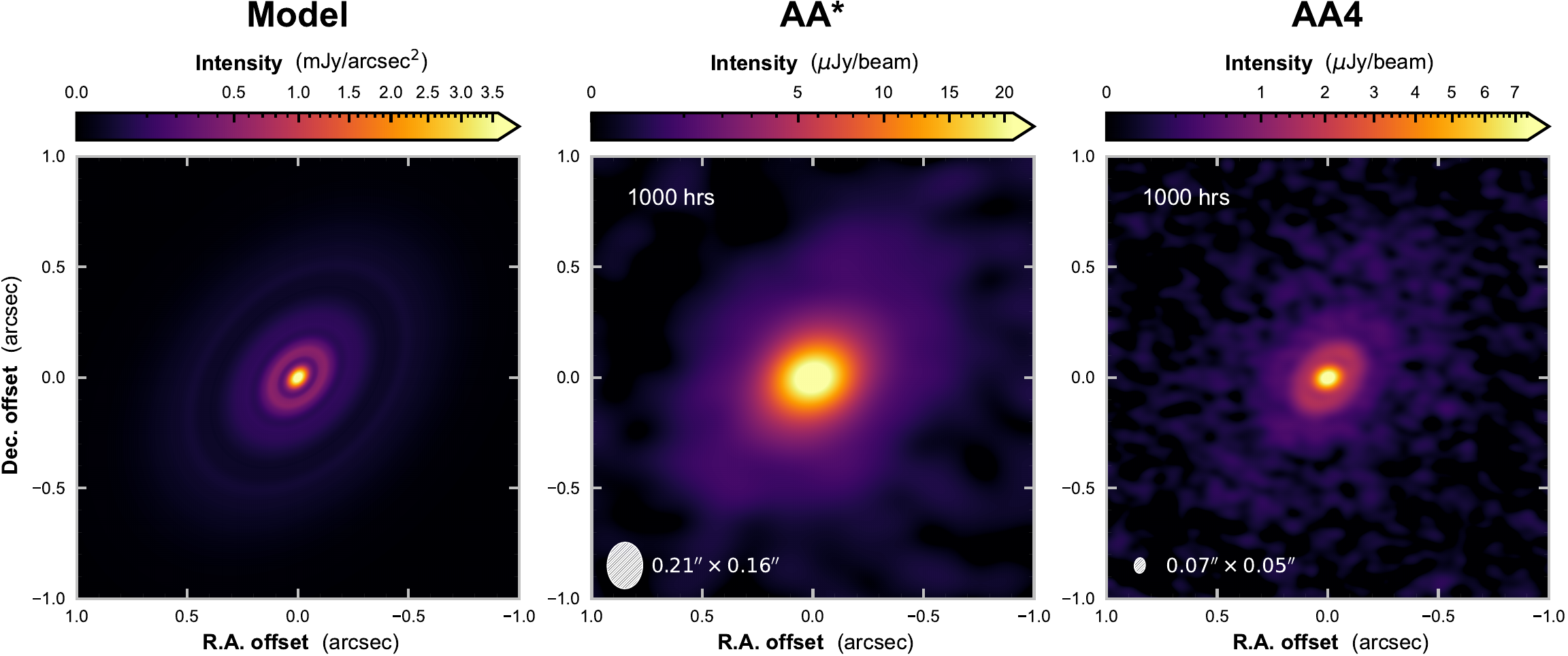}
\caption{HL Tau through the eyes of SKA-Mid: Predicted continuum images of the HL Tau disk at $12.5$~GHz (Band 5b) with AA* (middle panel) and AA4 (rightmost panel).  
The synthesized beams shown in the bottom left corners are generated from $uv$-coverage simulations, and weighted with a Briggs robust parameter of $-1$. 
Dish SKA008 is excluded from AA*. 
The achieved sensitivity is $0.115$~$\mu$Jy~beam$^{-1}$ and $0.049$~$\mu$Jy~beam$^{-1}$ by AA* and AA4, respectively. 
The leftmost panel shows the underlying model in $\mu$Jy~arcsec$^{-2}$. 
}\label{fig:hltau-AA*-AA4}
\end{figure}

In Figure \ref{fig:hltau-AA*-AA4}, we present simulated continuum images of the HL Tau disk with AA* (79 15-m antennas) and AA4 (133 15-m antennas), using 
the longest integration time 
shown in Figure \ref{fig:hltau-AA*-AA4-radialprofile}.  
The underlying sky model (presented in the left panel Fig \ref{fig:hltau-AA*-AA4}) is generated by azimuthally sweeping the model radial intensity profile and projecting it onto the sky with an inclination of $46.7^{\circ}$ and position angle of $138.0^{\circ}$ at a distance of 147~pc \citep{Carrasco-Gonzalez2019}. 
The sky model is convolved with a 2D Gaussian kernel representing the synthesized beam 
and correlated noise is added with an rms provided by the Sensitivity Calculator. 
We measure a flux density of approximately $85$~$\mu$Jy within a $0.9''\times0.6''$ elliptical aperture in both images.

\subsection{The SKA-Mid View of Other Structured Disks}\label{subsec:6-structured-disks-with-AA4}

Moving beyond a single benchmark object, we extend our SKA-Mid predictions to a broader sample of structured disk systems ---J1608, DL~Tau, V1094~Sco, PDS~70, HD~163296, and HD~142666--- that exhibit a range of transition and ring/gap morphologies, and some systems believed have evidence on host directly imaged accreting protoplanets, like PDS 70 \citep{Keppler2018,wagner2018-pds70, haffert2019-pds70} and WISPIT 2 \citep{Close-etal.2025WISPIT}. Here again our models can be anchored by high sensitivity, high angular resolution millimetre wavelength observations with ALMA. 
\citet{Villenave2025} conducted a comprehensive study to reproduce millimetre observations of structured disks with ALMA using the Monte Carlo radiative transfer code \texttt{mcfost} \citep{Pinte-etal.2006,Pinte-etal.2009}. 
The models consider axisymmetric disks with a single dust grain size distribution, e.g.\ $n(a) \propto a^{-3.5}$, with dust grain sizes $0.005\,{\rm\mu m} < a < 3\,{\rm mm}$ \citep{MRN1977}.  A gas-to-dust ratio of 100 is assumed to be constant across the disk, with the dust mass set to reproduce the total flux of the ALMA images. 
A Gaussian vertical density profile is assumed for the gas, with differential vertical dust settling set by the turbulent $\alpha$ parameter following \cite{Fromang_Nelson_2009}.  An iterative procedure was used to change the radial surface density of the model until this correctly reproduced the major axis brightness profile of the ALMA observations at either 0.8 or 1.3\,mm \citep[see also][]{Pinte2016, Pizzati_2023}. 

\begin{figure}
\centering
\includegraphics[width = 1\textwidth]{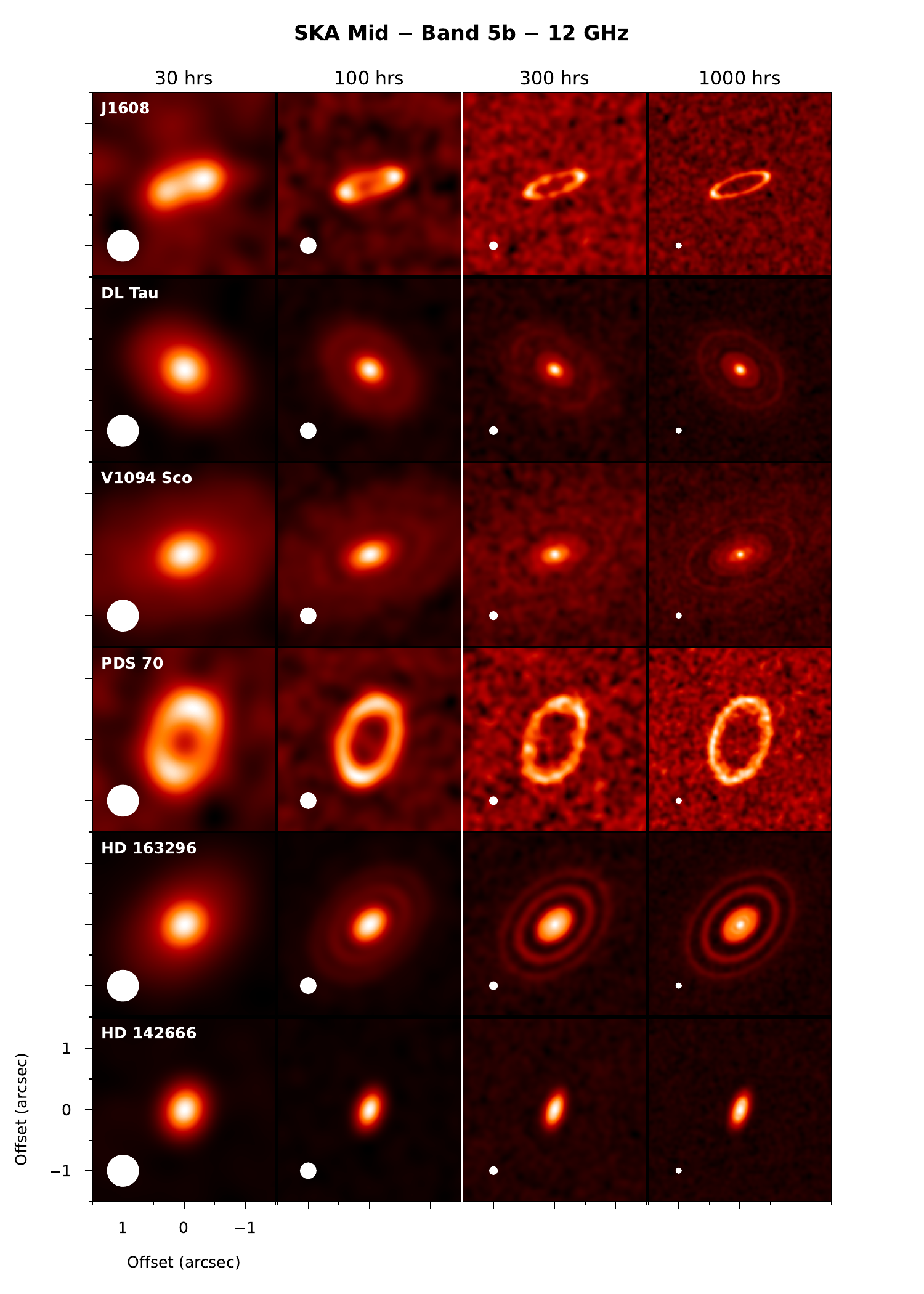}
\caption{
Simulated SKA Mid AA4 12.5\, GHz observations of six protoplanetary disks with improving angular resolution
($\ang{;;0.5}$, $\ang{;;0.25}$, $\ang{;;0.12}$ and $\ang{;;0.08}$) and increasing integration time (30, 100, 300 and 1000 hrs).  Each panel is peak normalised and shown with a power law stretch. 
}
\label{fig:ska_gallery}
\end{figure}

We have extended these radiative transfer calculations to predict the appearance of six of these disks at observing frequencies appropriate for SKA Mid's Band 5b.  We produce model images at a frequency of 12.5\,GHz (or wavelength of 2.4\,cm) using the model set-up described above. In all cases, we assume that the dust is vertically settled corresponding to $\alpha = 2\times10^{-3}$ but that the large and small grains are radially co-spatial.  This implies no substantial radial migration of large grains has taken place, which is an appropriate assumption for highly structured disks \citep[see, e.g.,][]{Pinilla2012b}.

We then simulate the observed appearance of the models under a variety of observing conditions based on information provided by the SKAO Sensitivity Calculator (version 2.1.1).  The radiative transfer models are first convolved with a Gaussian kernel of size and orientation corresponding to the synthesised beam for the requested observing conditions (assumed to lie at local zenith).  This is then added to a noise field sampled from a normal distribution (convolved with the same synthesised beam) and scaled to produce an image with an rms matching the weighted continuum sensitivity reported by the calculator (assuming a central frequency of 12\,GHz and bandwidth of 5GHz).  
We have verified that the results of this image plane technique are comparable to those obtained from considering a full pipeline of model visibility calculation and subsequent deconvolution \citep[e.g.][]{Ilee-etal.2020}, while also having the advantage of being computationally much faster.  This similarity is primarily due to the excellent $uv$-coverage of the AA4 configuration when considering the aperture synthesis of longer ($\gtrsim$12\,hr) integration times.  We explore a range of Briggs robust and $uv$-taper values that allow us to to push the synthesised beam size down with increasing integration time.  The results of this process are shown in Figure \ref{fig:ska_gallery}.

It is important to note that, by their design, the simulated SKA Mid observations of structured disks presented here do not include any emission from other mechanisms that may operate at SKA Mid frequencies, such as those from ionised gas.  Such emission mechanisms are discussed in detail by Guidi et al. in this volume, but some are likely to be bright and compact.  As such, they may introduce limitations on the achievable dynamic range for SKA Mid observations of dust continuum emission in disks.  Nevertheless, multi-frequency high angular resolution observations will be required to disentangle emission from these competing mechanisms, which SKA Mid will provide.


\subsection{Pushing the data to the limit: Analysing in visibility space}\label{subsec:visibility-analyses}



While facilities such as ALMA regularly provide high fidelity images of protoplanetary discs, there has been a renewed interest in modelling observations of disks directly in the native visibility plane.  Such approaches offer a number of advantages over image plane deconvolution techniques.  These include preserving the native spatial resolution and noise properties of the data, whilst also suppressing image reconstruction artefacts such as sidelobes and beam smearing.  \citet{Tazzari2018} introduced the \texttt{GALARIO} code that uses GPU acceleration to speed up the computation of synthetic visibilities.  \texttt{GALARIO} has been used to provide robust constraints on disk geometry, surface density and temperature profiles using analytic (e.g. self-similar) disk models applied to a wide range of observations of both individual disks \citep[e.g.][]{Clarke2018,Pinilla2021,Michel2022} and population-level studies of star forming regions such as Taurus, Orion, Ophiuchus, Lupus and Upper Scorpius \citep[see][]{Long2018,Sheehan2020,Tazzari2021,Vioque2025}.  Building on these foundations, \citet{Jennings2020} presented the \texttt{frank} code which uses a fast Gaussian process to directly fit observed visibilities non-parametrically and subsequently reconstruct one-dimensional radial surface brightness profiles for disks.  Using this method, \texttt{frank} is able to achieve sub-beam resolution when compared to what would be possible with CLEAN-based imaging methods, and has been used to reveal disk substructures such as rings, gaps and asymmetries on scales as small as a few au \citep[e.g.][]{Jennings2022a,Jennings2022b,Huang2024,Ribas2024}.

The application of visibility modelling to SKA Mid observations of protoplanetary disks would have many advantages.  \citet{Ilee-etal.2020} demonstrated that by conducting analysis of synthetic SKA Mid visibilities, it is possible to recover structural detail in disks that would only be observable using image-plane analysis using integration times 1-2 orders of magnitude longer.  Such improvements in efficiency, along with access to information on sub-beam scales, will be a powerful predictive tool as the Mid array evolves through each subsequent array assembly and operational mode.  However, there are also potential downsides.  The sheer volume of raw visibility data from SKA Mid will be many orders of magnitude larger than what is routinely delivered from facilities such as ALMA.  It is not clear how current visibility analysis tools would scale to such large datasets, or whether approaches based on alternative methods may be required. We therefore strongly advocate for access to calibrated visibilities during SKA Mid operations to enable the community to plan for the expansion of existing tools to exploit their analysis, ultimately enabling the highest scientific return from the observations.

\section{Discussion and Summary}

%

High angular resolution, high sensitivity observations at cm wavelengths are a crucial frontier in protoplanetary disk science. It is becoming increasingly clear that observations at mm wavelengths alone cannot resolve long-standing issues in planet formation.

When, exactly, planet formation begins remains a fundamental open question, tightly linked to the onset of observable substructure within disks. 
ALMA has revealed that rings, gaps, and other substructures (e.g. Fig. \ref{fig: substructures}) are common in Class II systems, 
but high optical depths at (sub-)mm wavelengths may be masking variations in the underlying mass distribution at earlier stages, obscuring them from our detection. 
Long-wavelength observations are essential to break this degeneracy: if the youngest disks are genuinely smooth, rather than simply optically thick, we should see this clearly at cm wavelengths. 
SKA-Mid offers a route to constrain the earliest emergence of substructure, and therefore the onset of planet formation itself.



A second, closely related question is whether the bright rings we now routinely observe at mm wavelengths are sites of grain growth and planetesimal formation (and therefore the \textit{birthplaces} of planets), or instead the signposts of planet-disk interaction (and therefore signposts of \textit{already-formed} planets). 
Distinguishing between these two scenarios requires assessing the grain size distribution within the substructures (not just across the disk as a whole),  
which demands spatially resolved spectral index measurements inside individual rings, and --critically-- parity in angular resolution across the spectral energy distribution. 
With SKA-Mid, we will finally achieve such parity from optical (HST) and infrared (JWST) through (sub-)mm (ALMA) to cm wavelengths. 
SKA-Mid will directly probe the largest grains and provide the much-needed lever arm to constrain the amount of grain growth inside these features.

Our simulated observations (\S\ref{sec:synthetic-predictions}) demonstrate SKA-Mid can resolve structured disks at cm wavelengths, but they also emphasize the need for high sensitivity -- which means focusing on the brightest disks and employing long integration times. 
As the community begins to strategize on priority targets and design programs, a balance should be struck between broad surveys and targeted deep imaging of the brightest and most extended disk systems. 
Both approaches are valuable, and indeed complementary: surveys to detect cm emission from many disks will reveal evolutionary trends in true dust mass measurements across different disk populations \citep[see also the Disk Demographics chapter by][]{Garufi-etal.2026}, while deep imaging will reveal the detailed morphology of individual exemplar disks. 



The synergy between SKA, ngVLA, and ALMA will offer an unprecedented, multi-wavelength view of protoplanetary disks \citep{Wu-etal.2024-SKA}. Together, they combine complementary strengths in resolution, sensitivity, and wavelength coverage across spatial scales, offering the best opportunity yet to disentangle thermal dust emission from free–free processes near the central star; a key step toward characterizing the physical conditions of inner disk regions. 


The SKAO's long-term vision foresees significant expansions that would further revolutionize this field. 
These upgrades, previously referred to as SKA2, could include a 10-fold increase in sensitivity in wavebands up to $15$~GHz, and an extension of maximum baselines to increase the angular resolution by a factor of $20$. 
In practical terms, this means the ability to detect even fainter, optically thin dust emission at angular resolutions approaching the mas scale (e.g. from more typical, lower-mass disks).


In summary, the Square Kilometre Array will bridge a long-standing observational gap, the centimeter-wavelength blind spot, by enabling us to directly trace the growth of solids within planet-forming disks. This capability will uncover hidden stages of early planet formation, providing crucial insights into how small dust grains evolve into the building blocks of planets.

Both SKA's high-sensitivity configurations (AA*) and its high-resolution mode (AA4) will be capable of detecting emission from centimeter-sized grains in disk substructures, precisely where planet formation seeds may be set. Beyond demographic studies (see Disk Demographics chapter), SKA will provide detailed views on individual systems, playing an essential role
in characterizing the distribution and properties of cm-sized dust
in protoplanetary discs.


\section{Appendix}\label{sec:appendix}

Dish SKA008 will be a part of SKA-Mid's composition starting from the AA2 array assembly (Science Verification), 
and will therefore also be a part of the subsequent AA* array assembly. 
SKA008 is located at the furthest end of one of AA4's spiral arms, isolated from the core of the AA* array layout. 
The AA* array assembly's maximum baseline of $108.0$~km is owed to SKA008's extended location; without this dish, AA*'s maximum baseline is $36.0$~km. 

\begin{figure}
\centering
\includegraphics[width=\textwidth]{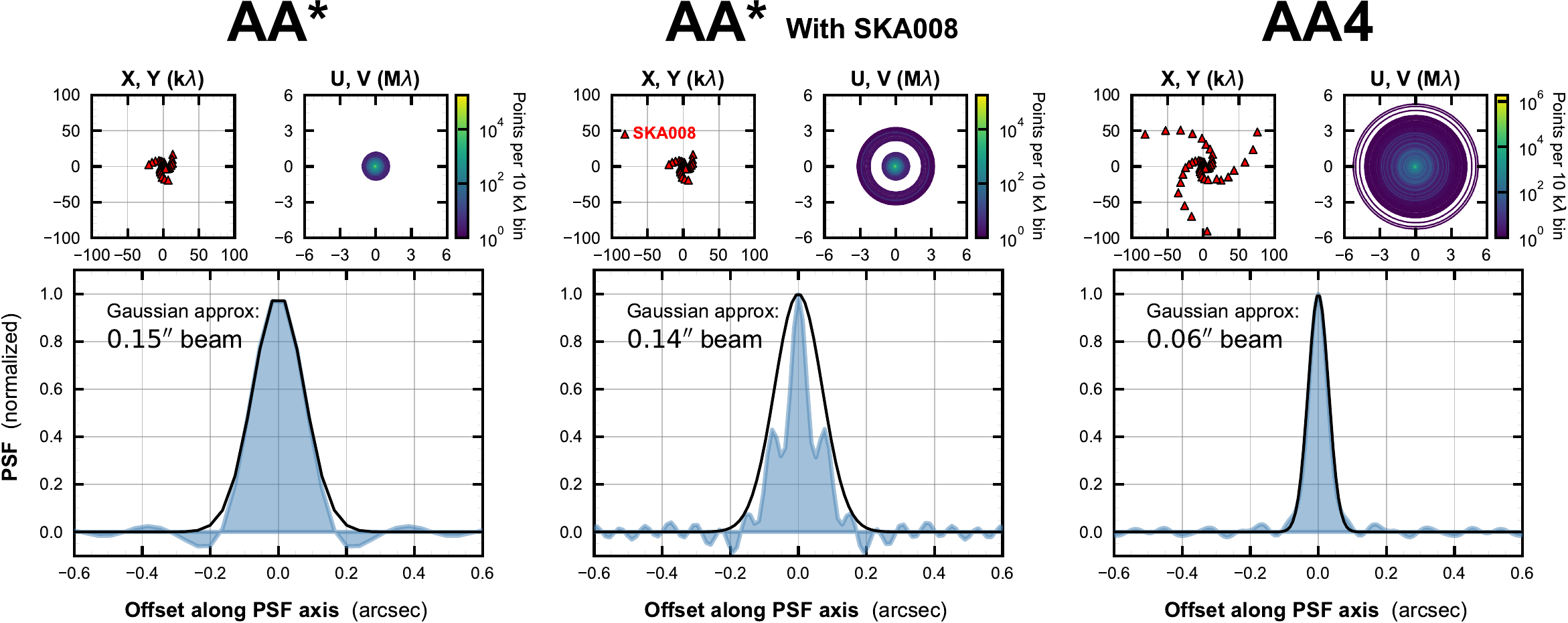}
\caption{Comparison of array layout, simulated $uv$ coverage, and PSF cross sections between array assemblies: AA* (left panels), AA* including dish SKA008 (middle panels) and AA4 (right panels). The $uv$-coverage simulations are performed with the SKAO Observing Support Tool \texttt{ska\_ost\_array\_config} python package. The pointing phase center is taken to be the celestial south pole ($-90.0^{\circ}$ declination), such that, after the set integration time of 12 hours, the PSF has become axisymmetric. Including dish SKA008 in AA* results in a highly non-Gaussian PSF with significant sidelobes. 
}\label{fig:PSF-AA*-AA4}
\end{figure}

In Figure \ref{fig:PSF-AA*-AA4}, we highlight SKA008's contribution to the $uv$-coverage of AA* observations, and the resulting impact on the point-spread function (PSF). For the simplicity of this demonstration, we adopt a pointing phase center of $-90.0^{\circ}$ declination and an integration of 12 hours, such that the $uv$ coverage and PSFs are axisymmetric.  
As shown in the left column of Figure \ref{fig:PSF-AA*-AA4}, the PSF of AA* \textit{without} SKA008 is approximately Gaussian. 
However, when dish SKA008 is included in AA* (as considered in the middle column of Figure \ref{fig:PSF-AA*-AA4}), 
it contributes an annulus of extended $uv$ coverage separated from that of the core. This in turn introduces non-Gaussianity and significant sidelobes within the AA* point-spread function. 
The right column of Figure \ref{fig:PSF-AA*-AA4} shows the PSF of AA4 is again approximately Gaussian, with its three spiral arms fully stationed. 

As described in \S\ref{sec:synthetic-predictions}, we omit dish SKA008 when simulating AA* observations, as our image-plane approach assumes a Gaussian PSF. 
The SKAO Observing Support Tool (\texttt{ska\_ost\_array\_config} python package) enables us to manually omit SKA008 from the AA* array layout when generating $uv$ coverage simulations. 
We stress that we adopt sensitivity estimates for AA* directly from the SKAO Sensitivity Calculator, which includes SKA008 in the estimate. We have not accounted for the discrepancy in the number of dishes (the difference between 79 and 80) nor the mis-matching beam sizes. 




\section*{Acknowledgements}
\textit{This research was supported by the funding from the National SKA Program of China under Grant No. 2025SKA0120100. Y.W. acknowledges the EACOA Fellowship awarded by the East Asia Core Observatories Association. S.P. acknowledges support from FONDECYT 1231663, ANID--Millennium Science Initiative Program NCN2024$\_$001 and ANID FIUF137139-USACH. J.D.I. acknowledges support from an STFC Ernest Rutherford Fellowship (ST/W004119/1). E.B. acknowledges support from the Italian Ministry for Universities and Research under the Italian Science Fund (FIS 2 Call – Ministerial Decree No. 1236 of 2023 August 1) grant FIS-2023-00170. G.B. acknowledges support from the grant PID2023-146675NB-I00 (MCI-AEI-FEDER, UE) and the grant CEX2024-001451-M funded by MICIU/AEI/10.13039/501100011033. Part of this research by M.N. was carried out at the Jet Propulsion Laboratory, California Institute of Technology, under a contract with the National Aeronautics and Space Administration (80NM0018D0004).}

\textit{We thank the referee for a thorough review and highly constructive suggestions, which significantly enhanced the quality of the manuscript. The authors acknowledge the assistance of artificial intelligence in improving the readability of the text.
}

\bibliographystyle{abbrvnat-maxbibnames4}
\bibliography{chapter} 

\end{CJK*}
\end{document}

%% file: journal-names.tex
\newcommand{\actaa}{Acta Astron.} 
\newcommand{\araa}{ARA\&A} 
\newcommand{\aar}{A\&ARv} 
\newcommand{\aapr}{A\&ARv} 
\newcommand{\ab}{Astrobiol.} 
\newcommand{\aj}{AJ} 
\newcommand{\apj}{ApJ} 
\newcommand{\apjl}{ApJL} 
\newcommand{\apjs}{ApJSS} 
\newcommand{\ao}{Appl. Opt.} 
\newcommand{\apss}{Astro. \& Space Sci.} 
\newcommand{\aap}{A\&A} 
\newcommand{\aaps}{A\&AS.} 
\newcommand{\baas}{Bull. Am. Astron. Soc.} 
\newcommand{\caa}{Chinese A\&A} 
\newcommand{\cjaa}{Chinese J. A\&A} 
\newcommand{\cqg}{Class. Quantum Gravity} 
\newcommand{\gal}{Galaxies} 
\newcommand{\gca}{Geo. Cosmo. Acta} 
\newcommand{\icarus}{Icarus} 
\newcommand{\jcap}{JCAP} 
\newcommand{\jgr}{J. Geophys. Res.} 
\newcommand{\jgrp}{J. Geophys. Res. Planets} 
\newcommand{\jqsrt}{J. Quant. Spectrosc. Radiat. Transf.} 
\newcommand{\memsai}{Mem. SAIt} 
\newcommand{\mnras}{MNRAS} 
\newcommand{\nat}{Nature} 
\newcommand{\nastro}{Nat. Astron.} 
\newcommand{\ncomms}{Nat. Commun.} 
\newcommand{\nphys}{Nat. Phys.} 
\newcommand{\na}{New Astron.} 
\newcommand{\nar}{New Astron. Rev.} 
\newcommand{\physrep}{Phys. Rep.} 
\newcommand{\pra}{Phys. Rev. A} 
\newcommand{\prb}{Phys. Rev. B} 
\newcommand{\prc}{Phys. Rev. C} 
\newcommand{\prd}{Phys. Rev. D} 
\newcommand{\pre}{Phys. Rev. E} 
\newcommand{\prx}{Phys. Rev. X} 
\newcommand{\prl}{Phys. Rev. Let.} 
\newcommand{\psj}{Planet. Sci. J.} 
\newcommand{\planss}{Planet. Space Sci.} 
\newcommand{\pnas}{Proc. Natl Acad. Sci. USA} 
\newcommand{\procspie}{Proc. SPIE} 
\newcommand{\pasa}{PASA} 
\newcommand{\pasj}{PASJ} 
\newcommand{\pasp}{PASP} 
\newcommand{\rmxaa}{RMXAA} 
\newcommand{\sci}{Science} 
\newcommand{\sciadv}{Sci. Adv.} 
\newcommand{\solphys}{Sol. Phys.} 
\newcommand{\sovast}{Soviet Ast.} 
\newcommand{\ssr}{Space Sci. Rev.} 
\newcommand{\uni}{Universe} 